\documentclass[12pt]{article}
\usepackage[dvips]{graphicx,color}
\usepackage{overcite,dcolumn,amsmath}

\newcommand{\GlobalFigScale}{.85}
\definecolor{rulecolor}{gray}{0.5}
\newsavebox{\HttpBox}
\begin{lrbox}{\HttpBox}
\footnotesize\verb+http://www.mpip-mainz.mpg.de/~yasp/ccteam/index-gruppe.html+
\end{lrbox}
\newsavebox{\EBox}
\begin{lrbox}{\EBox}
\footnotesize\verb+reith@mpip-mainz.mpg.de+
\end{lrbox}

\newcommand{\caplist}[1]{}

\begin{document}
\title{\Large How does the chain extension of poly (acrylic acid) scale
  in aqueous solution? A combined study with light scattering and computer
  simulation} 
\author{Dirk Reith\footnote{corresponding author, Electronic
    Mail: \usebox{\EBox}}, Beate M\"uller, Florian
  M\"uller-Plathe\footnote{Homepage: \usebox{\HttpBox}},
  ~and Simone Wiegand\\
  \small Max-Planck-Institut f\"ur Polymerforschung, D-55128 Mainz, Germany}
\date{For submission to \textsl{Macromolecules}\\\today}
\maketitle
%
%
%
\begin{abstract}
  This work adresses the question of the scaling behaviour of polyelectrolytes
  in solution for a realistic prototype: We show results of a combined
  experimental (light scattering) and theoretical (computer simulations)
  investigation of structural properties of poly (acrylic acid) (PAA).
  Experimentally, we determined the molecular weight ($M_{W}$) and the
  hydrodynamic radius ($R_{H}$) by static light scattering for six different
  PAA samples in aqueous NaCl-containing solution ($0.1-1$ mol/L) of
  polydispersity $D_P$ between $1.5$ and $1.8$. On the computational side,
  three different variants of a newly developed mesoscopic force field for PAA
  were employed to determine $R_{H}$ for monodisperse systems of the same
  $M_{W}$ as in the experiments. The force field effectively incorporates
  atomistic information and one coarse-grained bead corresponds to one PAA
  monomer.  We find that $R_{H}$ matches with the experimental data for all
  investigated samples.  The effective scaling exponent for $R_{H}$ is found
  to be around $0.55$, which is well below its asymptotic value for good
  solvents. Additionally, data for the radius of gyration ($R_{G}$) are
  presented.
\end{abstract}

%
%
%
\section{Introduction}
%
%
%
Poly (acrylic acid) (PAA) is a water-soluble polyelectrolyte (PE). It is
important not only in industrial applications, e.g. flocculants or
superabsorbers~\cite{cloizeaux90}. Because of its relatively simple chemical
repeat unit, it is also a prototype PE model for scientific investigations.
Surprisingly, there seems to be no (accurate) data on some solution properties
of PAA chains. Quantities which characterize the solution behaviour of
isolated polymer chains are the radius of gyration $R_G$ and the hydrodynamic
radius $R_H$. It is the main purpose of this letter to provide reliable
values for PAA of different chainlengths.

Two approaches are being used. Experimentally, we determine the size of a
polymer coil in solution by dynamic ($R_H$) or static ($R_G$) light
scattering. The measurements were performed using narrow fractions of
radically polymerized PAA in order to, on the one hand, minimize effects of
polydispersity and to, on the other hand, avoid possible problems of PAA
aggregation due to hydrophobic initiators used in anionic
polymerization~\cite{Huber-priv}. These data are augmented by results from
computer simulations. Simulations have the principal advantage that
macroscopic observations can be understood in terms of a microscopic model.
We will use them here to investigate the scaling of the polymer
also at length scales shorter than the overall coil size ($R_G$ or $R_H$)
which is accessible by the light-scattering experiment. This is
the second aim of this work.

The experimentally relevant molecular weights are far beyond what can be
simulated with an atomistic model. They are, however, accessible with suitably
simplified or "coarse-grained" (CG) models. Care has to be taken that the CG
model retains sufficient information about the chemical nature of the polymer.
A generic bead-spring model might be good enough for theoretical scaling
relations~\cite{degennes:79,doi:86,barrat96a}, but will not give realistic absolute
values for $R_H$ of specific polymers like PAA. A number of systematic
coarse-graining procedures have been
described~\cite{flory69,ryckaert75,carmesin88,paul91,hoogerbrugge:92,forrest95,rapold95,zimmer96,ahlrichs98,eilhard99,akkermans01},
but none has been tried for specific PEs in solution. We have recently developed an
automatic coarse-graining method which works by a two-step process: Firstly,
an atomistic simulation of a PAA oligomer in water is performed. Secondly, the
CG model is parametrized to reproduce the PAA structure~\cite{reith00s}. To
validate this method and to try out technical variants on it is the third
objective of the letter.

%
%
%
\section{Methods\label{methods}}
%
%
%

%
%
%
\subsection{Experimental}
\subsubsection{Static and Dynamic Light Scattering}

For simultaneous static and dynamic light scattering measurements a commercial
instrument (ALV-5000) with a krypton ion laser operating at a wavelength of
647 nm and an Avalanche Diode (EG\&G) as detector was used \cite{Vanhee-1996}.
Static measurements were performed at scattering angles of
$30^{\circ}-150^{\circ}$ in $10^{\circ}$ steps in the concentration range of
$0.5-5$ g/L. For the high molecular weights the highest concentration was only
$3.5$~g/L in order to remain in the dilute regime. Weight average molar masses
$M_{W}$ and radii of gyration $R_{G}$ were obtained by Zimm extrapolation
using the Rayleigh ratio $RR=1.270~{\mathrm{cm}}^{-1}$. The refractive index
increments $dn/dc$ are listed in Table~\ref{tab:expdata}, as measured at
$\lambda = 663$ nm using a scanning Michelson interferometer
~\cite{becker-1995}.

Dynamic measurements were performed over an angle range from
$60^{\circ}-150^{\circ}$ in $30^{\circ}$ steps. The intensity autocorrelation
function of the scattered intensity was converted to the scattered electric
field and analyzed using the program CONTIN by S.Provencher
\cite{Provencher-1982}. Hydrodynamic radii $R_{H}$ were obtained via the
Stokes-Einstein relation, where the apparent diffusion coefficient $D=\Gamma
/ q^2$ was calculated from the inverse relaxation time $\Gamma$ and the
absolute value $q$ of the scattering vector.\cite{Berne-and-Pecora-1964}
Viscosity corrections due to the NaCl were considered to be small and not
taken into account. The hydrodynamic radius $R_H$ was measured repeatedly.
In Figure~\ref{fig:exp_checks} the average values and error bars representing
one standard deviation are plotted.

\subsubsection{Sample Preparation}
The PAA samples (Polymer Standard Service) with molar masses of
$M_{W}=18000-296600$ g/mol were synthesized by radical polymerization and
fractionated with polydispersities $D_P=M_W/M_N$ between 1.5 and 1.8 (with
$M_N$ being the number average molecular weight). All samples were dissolved
in deionized water (Millipore) with $1M$ NaCl.  The $pH$ was determined for
two samples ($M_{W}=81800$ and $296600$) at two different concentrations ($1$
g/L and $5$ g/L) and found to lie between $6-8$. After further purification,
we performed additional measurements by dialysis using a regenerated cellulose
membran (MWCO2000). The dialysis process lasted 40 hours and the deionized
water (Millipore) was replaced four times.  Afterwards, the samples were
freeze-dried for 5 hours and finally dried under vaccuum for 60 hours at $T =
303$ K, in order to avoid fluctuations of the PAA concentration by adsorbed
water. The refractive index increments for the dialyzed and dried samples
remained unchanged within the statistical errors.  All solutions were filtered
through a $0.22$ $\mu$m Millex-GS filter (Millipore) to remove dust particles.

\subsection{Computational\label{concepts}}
\subsubsection{The Coarse-grained Force Field for PAA}
Coarse-grained potential energy functions for polymers have to incorporate not
only energetic, i.e.\ local aspects of the underlying microscopic model.  They
also have to account for entropic contributions from the neglected
conformational degrees of freedom of the chain.  Therefore, we utilized
structural information obtained from fully atomistic simulations as target
functions to construct our CG force field. In particular, we used the intra-
and inter-chain radial distribution functions which, being distributions
derived from an ensemble, contain the desired entropic information. This is a
so-called inverse problem: find an interparticle potential which reproduces a
given radial distribution function (RDF) or set of RDFs.  For the fitting
procedure, we applied an automatic optimization algorithm (\textsl{simplex})
which was originally implemented for the development of atomistic force
fields~\cite{faller99} and tested for liquids.\cite{meyer00} Our force field
is based on atomistic simulation data by Biermann et al.\cite{biermann01s}.
They studied one fully deprotonated, atactic oligomer of 23 monomers with 23
Na$^+$ counterions in 3684 water molecules (simple point charge (SPC)
model~\cite{berendsen:81}) at ambient conditions.  The system represents a
diluted ($\approx 2.3$wt$\%$) PAA solution. From this simulation, structural
information like the distributions of bond angles or RDFs between monomers
were extracted.  We mapped this system to the mesoscale by replacing each
repeat unit (i.e.\ each monomer) by one bead, either at the monomers center of
mass or at the backbone carbon bearing the carboxyl group.  The PAA mesoscale
force field includes bonded and non-bonded interactions.  To both contribute
several terms~\cite{jensen99}. They were parametrised by systematically
varying the interactions until the structure of the atomistic model was
reproduced. This also allowed us to omit all explicit water molecules and
counter ions. Their effect on the PAA chain conformation is, however,
implicitly present in the model.  In effect, a system of roughly $10000$ Atoms
could thus be reduced to a system which consists of only $23$ "superatoms".

For the sake of gaining experience with CG force fields, we have
coarse-grained PAA in three different ways. Potential $I$ is pieced together
from various analytic functions. It contains attractive as well as repulsive
intermolecular interactions. Coulomb forces were only implicitly taken into
account (they were present in the parent atomistic simulation). That is
because the Debye length (characterizing the relative importance of the
electrostatic energy compared to thermal energy) of the system is close to
that of water, so that electrostatic interactions can be treated as being
short-ranged. For details, see ref 17. Potential $II$ is similar in
construction, but uses another location of the CG bead: While Potential $I$
placed it at the center of mass of a monomer, Potential $II$ employed the
backbone carbon parent to the COO$^{-}$-group~\cite{reith01phd}. As example
for the achieved convergence, the RDF (first neighbours excluded) is shown in
Figure~\ref{fig:rdfs}a.  Potential $III$ uses again the center-of-mass of the
PAA monomer, but was optimized with a fully numerical instead of a piecewise
analytical potential energy function for the non-bonded
interactions~\cite{reith01phd}. It allows us to apply a new self-consistent
optimization algorithm which was recently adapted by us.~\cite{puetzreith01}
The optimization then yields an almost perfect match of the RDF.  Target RDFs
(exluding first and second neighbours, as defined in Ref.\cite{reith00s}) and
the results of the optimization are shown in Figure~\ref{fig:rdfs}b. The
parameters of force fields $I$ and $II$ are presented in
Table~\ref{tab:parms}. The non-bonded part of Potential $III$ exists in
numerical form only~\cite{reith01phd} (not shown here).

\subsubsection{Technical Simulation Details}
Both Brownian Dynamics (BD) and Monte Carlo (MC) programs were used to carry
out the PAA simulations. We simulated at $T=333.15$ K, which corresponds to the 
temperature of the atomistic simulations. The Langevin equations of
motion were integrated by the velocity Verlet algorithm with a time step
$\Delta t = 0.01\tau$,\cite{All87} and a friction constant $\gamma = 0.5
\tau^{-1}$.\cite{grest86} Pivot MC calculations were necessary to simulate
systems of more than $1000$ monomers as the equilibration is much faster. We
carried out $10^5$ accepted equilibration moves before a production run of
$10^6$ accepted moves started.

%
%
%
\section{Results and Discussion}
%
%
%
The hydrodynamic radii of six different PAA samples with molecular weights in
the range from $18100$ to $296600$ g/mol were measured (see
Table~\ref{tab:expdata}). For four samples, the molar masses $M_{W}$ and the
radii of gyration $R_{\mathrm{G}}$ were measured in 1M NaCl solution. (For the
samples of lower molecular weight an accurate determination of $R_G$ was not
possible, so they are reported as specified by the supplier.) Additionally,
the ratio $R_G/R_H$ and the refractive index increment $\partial n/\partial c$
are shown in Table~\ref{tab:expdata}.  In all PAA solutions, the salt
content was so high that the electrostatic interactions were effectively
screened.  Therefore, $R_G$ could be obtained from the static scattering
experiment by Zimm extrapolation to zero angle and zero polymer concentration
$c_{p}$ like for uncharged polymers.  The dynamic measurements in $1$M
NaCl solution yielded single-exponential autocorrelation functions, since we
performed our measurements in the "ordinary regime" where
$c^{\mathrm{m}}/c_{\mathrm{s}}^{\mathrm{m}}<1$, with $c^{\mathrm{m}}$ and
$c_{\mathrm{s}}^{\mathrm{m}}$ being the molar concentration of monomer units
and of salt, respectively.\cite{Dautzenberg-1994} No polyelectrolyte slow mode
due to a cooperative diffusion behavior was observed. Furthermore, the
diffusion coefficients do not show any significant angular dependence.

For two samples with molecular weigths of $M_{W}=119800$ g/mol and
$M_{W}=204000$ g/mol, we varied the NaCl concentration between
$0.1-1$~mol/L at a fixed polymer concentration of $1$ g/L, as shown in
Figure~\ref{fig:exp_checks}a. No dependence of the diffusion coefficient or
$R_H$ is found within this range.  Figure~\ref{fig:exp_checks}b shows a weak
decrease of $R_H$ with polymer concentration in the range of~$0-5$~g/L for
two molecular weights $M_{W}$ at fixed NaCl concentration of $1$ M.  For
further comparisons we use the values extrapolated to zero concentration.
Thus the static as well as dynamic light scattering experiments represent
measurements of single polylelectrolyte molecules. The hydrodynamic radii are
plotted in Figure~\ref{fig:scaling} as a function of molecular mass.  We find
a scaling exponent of $\nu=0.56\pm0.02$. This scaling exponent
compares well with CG simulation results (below).

In Table~\ref{tab:simdata}, the corresponding theoretical data is presented
for $13$ different chain lengths (being single chains, the simulated polymers
are monodisperse).  They include all the experimental ones when converting the
mean molar weights $M_W$ into the degree of polymerization. We plotted them in
Figure~\ref{fig:scaling}a, too. Over the whole range of measured samples, the
coincidence is excellent. This is especially remarkable if one considers the
fact that the simulation model was developed with a PAA chain consisting of
only $23$ monomers.

In order to check the transferability of parameters between oligomer chains of
different lengths, we executed new atomistic simulations of shorter chains
with the same force field as used by Biermann~et~al. in their
work.~\cite{biermann01s} The PAA was dissolved in $4000$ water molecules
(concentration $c_{p} \approx 1$ g/L) and the total simulation time was $6$
ns.  The concentration of sodium counterions was determined to be around
$0.44$ mol/L for all atomistic simulations, which lies in the middle of the
experimentally tested ones. A "downward" transferability could be obtained
for both samples. The CG model developed for a $23$-monomer chain, faithfully
reproduces also the $R_{H}$ and $R_{G}$ of the atomistic $8$- and $12$-mers,
cf.\ Table~\ref{tab:simdata}a. It is also interesting to investigate the
stability of $R_H$ for the various simulation models $I$, $II$ and $III$. As
shown in Table~\ref{tab:simdata}b, the coincidence is good for all chain
lengths.  This means that our CG mapping is not unique.  Structural properties
can be determined from any of the above model variants. That might be of use,
if one is interested in properties pertaining to the center-of-mass or of the
CH-backbone carbon.

In the following, we focus our attention on the original force field $I$.  The
simulated $R_{H}$ for chains longer than $100$ monomers was fitted by a power
law $R_{H} \propto M_{W}^{\nu}$. The resulting scaling exponent $\nu$ was
$0.55\pm0.02$. If we restrict the fit of the experimental data to the
simulated chain lengths, we obtain the same exponent. Experiments and
simulation yield, hence, very similar results. It can be concluded that the
salt concentration (used in experiments) or sodium counterion concentration
(in the parent atomistic simulations) effectively screen the charges of the
PAA chain.  Thus, the Coulomb potential of PAA is really reduced to an
effective short-ranged interaction. For small distances, though, the charges
stiffen the chain. This behaviour can be understood in the framework of
scaling theory. For a Gaussian chain (model of a linear
\textsl{fully-flexible} polymer) in a good solvent, self-avoiding walk (SAW)
statistics applies. It yields a scaling exponent of $\nu=0.588$ in the limit
of infinitely long chains~\cite{doi:86}. This scaling factor is universal,
i.e.\ it is valid for every size-characterizing function, e.g.\ $R_{G}$ or
$R_{H}$.  The structur factor $S(q)$, however, is an even better function to
investigate. That is, because it scales as $S(q) \propto q^{-1/\nu}$, which we
can try to fit for any length scale up to the total size of the polymer. So,
local deviations from the global behaviour, as measured by the experiment, can
be observed.  Figure~\ref{fig:scaling}b shows the structure factor of the
coarse-grained BD simulation of a PAA $460$-mer. Apart from noise in the
region $q>10$, three different regimes can be distinguished: On very short
distances (nearest neighbours), the slope equals $-1.65=-1/\nu$, i.e.\ 
corresponding to a "local" scaling exponent of $\nu=0.61$. This is remarkably
close to, but larger than the value for a SAW ($0.588$).  Compared to a
Gaussian chain, this is due to the stiffening bond angle potential of the
chain (in the CG picture). Next comes the region in which torsions and
non-bonded interactions become important. They introduce even more stiffness,
resulting in a large exponent of $\nu=0.82$.  Finally, the long-range
behaviour sets in ($\nu=0.56$), almost yielding the obtained total scaling
exponent for $R_{H}$.  So we can conclude that PAA globally behaves like a
fully-flexible chain without charges. These results are very similar for any
other large-enough PAA sample. We note also that the original target RDF of
the 23-mer is reproduced by any $23$-monomer sub-strand of the $460$-mer (data
not shown).  For $R_{G}$, scaling similar to that of $R_{H}$ is, as expected,
obtained in the limit of sufficiently long chains ($> 100$ monomers). In
Figure~\ref{fig:scaling}a, this can be seen by examining the ratio $R_G/R_H$:
a plateau region is reached for long chains, with a value of around $1.5$.
This is the value predicted for Gaussian chains. A closer look, however,
reveals a monotonically increasing ratio (cf.\ Table~\ref{tab:simdata}),
exceeding the value of $1.5$. The reason for this behaviour are the
corrections to scaling due to the good solvent statistics of the chain, which
have been known for a long time~\cite{batoulis89}.  For $R_{H}$, this
motivated the empirical indroduction of an effective exponent $\nu_{eff}
\approx 0.55$,\cite{adam76} which exactly equals our findings. A detailed
analysis of these corrections, whose asymptotic value is found to be
$R_{G}/R_{H} \approx 1.61$, will be presented elsewhere~\cite{duenweg01}.

\section*{Acknowledgements}
Oliver Hahn (\verb+md_spherical+) and Mathias P\"utz (\verb+prism+) are
gratefully acknowledged for making available the code of their simulation
programs. We are indebted to Franziska Gr\"ohn for her contributions to our
work. We also acknowledge fruitful discussions with Kurt Kremer, Manfred
Schmidt, Klaus Huber and Thorsten Hofe, whom we also thank for free PAA
samples and complementary measurements with an online light scattering
detector.

%
%

\clearpage


%
%
%
%
%
%
\clearpage

%
%
\thispagestyle{empty}
\begin{table}[htbp]
\vspace*{-1.5cm}
  \begin{center}
    \caption{Parameters for our various coarse-grained force fields. For
      details, see refs 17 and 27. }
    \begin{tabular}{lcccc}
&       &            &          & \\
(a) bond angle parameters &&&       & \\
&       &            &          & \\
& Peak \# & position & standard deviation & peak height\\
&       & [$^{o}$]   &  [$^{o}$] & [height of peak 2]\\
&       &            &          &  \\ \hline\hline
center of mass & & & &\\
force fields $I$,$III$& & & &\\
& 1 & 88.0 & 6.8 & 1.72\\
& 2 & 116.7 & 7.4 & 1.00\\[10pt]\hline
backbone carbon & & & &\\
force field $II$& & & &\\
& 1 & 127.4 & 8.5 & 1.15\\
& 2 & 156.4 & 8.9 & 1.00\\[10pt]
    \end{tabular}\
    \begin{tabular}{lcccccccc}
& & & & & & & & \\
(b) non-bonded parameters& & & & & & & & \\
& & & & & & & & \\
& $\sigma_{1}$ & $\varepsilon_{1}$ & $\sigma_{2}$ & $\varepsilon_{2}$ %
& $\sigma_{3}$ & $\varepsilon_{3}$ & $\sigma_{4}=r_{cut}$ & $\varepsilon_{4}$ %
\\
& & & & & & & & \\ \hline\hline
original force field & & & & & & &\\
(center of mass) $I$& & & & & & &\\
& 0.496 & 11.30 & 0.559 & 0.35 & 0.775 & 0.49 & 1.3 & 0.05\\[10pt] \hline
backbone carbon& & & & & & &\\
force field $II$& & & & & & &\\
& 0.496 & 11.30 & 0.565 & 1.20 & 0.802 & 0.70 & 1.3 & 0.05\\[10pt]
    \end{tabular}
    \label{tab:parms}
  \end{center}
\end{table}

\newpage

\thispagestyle{empty}
\begin{table}[htbp]
  \begin{center}
    \caption{Experimental hydrodynamic radii $R_{H}$ and radii of gyration
      $R_{G}$ obtained by dynamic and static light scattering experiments for
      poly (acrylic acid) of different molecular weights $M_{W}$.}
    \label{tab:expdata}
    \begin{tabular}{lcccc}
&&&\\\hline\hline
$M_W$ [g/mol] & $R_{G}$ [nm] &  $R_{H}$ [nm] & $R_{G}/R_{H}$ & $\partial n/\partial c$\\\hline\hline 
18100$^1$  (193-mer)  & -    & 3.7  & - & -\\
36900$^1$  (393-mer) & -    & 5.3  & - & -\\
81800$^2$  (870-mer) & 12.9 & 8.9  & 1.45 & 0.1360\\
119800$^2$ (1287-mer) & 14.9 & 10.4 & 1.44 & 0.1452\\
204000$^2$ (2170-mer) & 22.7 & 15.0 & 1.51 & 0.1472\\
296600$^2$ (3155-mer) & 23.8 & 16.6 & 1.46 & 0.1310\\\hline\hline
    \end{tabular}
  \end{center}
  $^1$: $M_{W}$ as specified by the supplier\\
  $^2$: $M_{W}$ determined by own static light scattering
\end{table}

\newpage

\thispagestyle{empty}
\begin{table}[htbp]
  \begin{center} 
    \caption{Radii of gyration $R_{G}$ and hydrodynamic radii $R_{H}$ for
      different-size poly (acrylic acid) as obtained by Brownian Dynamics
      (marked '1') and Monte-Carlo (marked '2') simulations. In
      the coarse-grained picture, potentials $I$ and $III$ are different
      versions of center-of-mass mapped force field, whereas potential $II$ is 
      centered at the backbone carbon of the chain.}
    \label{tab:simdata}
    \begin{tabular}{lccc}
&&&\\\hline\hline
Chainlength & $R_{G}$ [nm] &  $R_{H}$ [nm] & $R_{G}/R_{H}$\\\hline\hline
&&&\\
(a) Atomistic Model&&&\\\hline
8-mer  & 0.55 & 0.87 & 0.63\\
12-mer & 0.77 & 0.91 & 0.85\\
23-mer & 1.28 & 1.19 & 1.08\\\hline\hline
&&&\\
(b) Coarse-Grained Model&&&\\\hline
Potential $I$ &&&\\
8-mer$^{1,2}$  & 0.57 & 0.89 & 0.64\\
12-mer$^1$ & 0.78 & 0.94 & 0.83\\
23-mer$^{1,2}$ & 1.25 & 1.18 & 1.06\\
46-mer$^2$  & 2.0 & 1.6 & 1.25\\
92-mer$^2$  & 3.1 & 2.3 & 1.35\\
193-mer$^1$ & 4.8  & 3.4 & 1.41 \\
393-mer$^1$ & 6.8  & 4.8 & 1.42 \\
460-mer$^1$ & 7.7  & 5.3 & 1.45 \\
460-mer$^2$ & 7.9  & 5.5 & 1.44 \\
852-mer$^1$ & 10.8 & 7.4 & 1.46 \\
855-mer$^2$ & 11.3 & 7.7 & 1.47 \\
1287-mer$^1$& 13.5 & 9.1 & 1.48 \\
2067-mer$^2$& 18.8 & 12.4& 1.52 \\
3155-mer$^2$& 24.1 & 15.7& 1.54 \\\hline
Potential $II$ &&&\\
23-mer$^1$  & 1.22 & 1.03 & 1.18\\
393-mer$^1$ & 6.9  & 4.8  & 1.43\\
852-mer$^1$ & 9.4  & 6.6  & 1.42 \\
1287-mer$^1$& 14.3 & 9.0  & 1.59 \\\hline
Potential $III$ &&&\\
23-mer$^1$  & 1.26 & 1.18 & 1.07\\
460-mer$^1$ & 8.6  & 5.9  & 1.46\\
852-mer$^1$ & 12.2 & 8.3  & 1.47 \\
1287-mer$^1$& 15.5 & 10.3 & 1.51 \\\hline\hline
    \end{tabular}
  \end{center}
\end{table}


%
%
%
%
%
%
\clearpage

\parindent0mm
\pagestyle{empty}

\begin{figure*}
 \[ \includegraphics[width=\GlobalFigScale\linewidth]{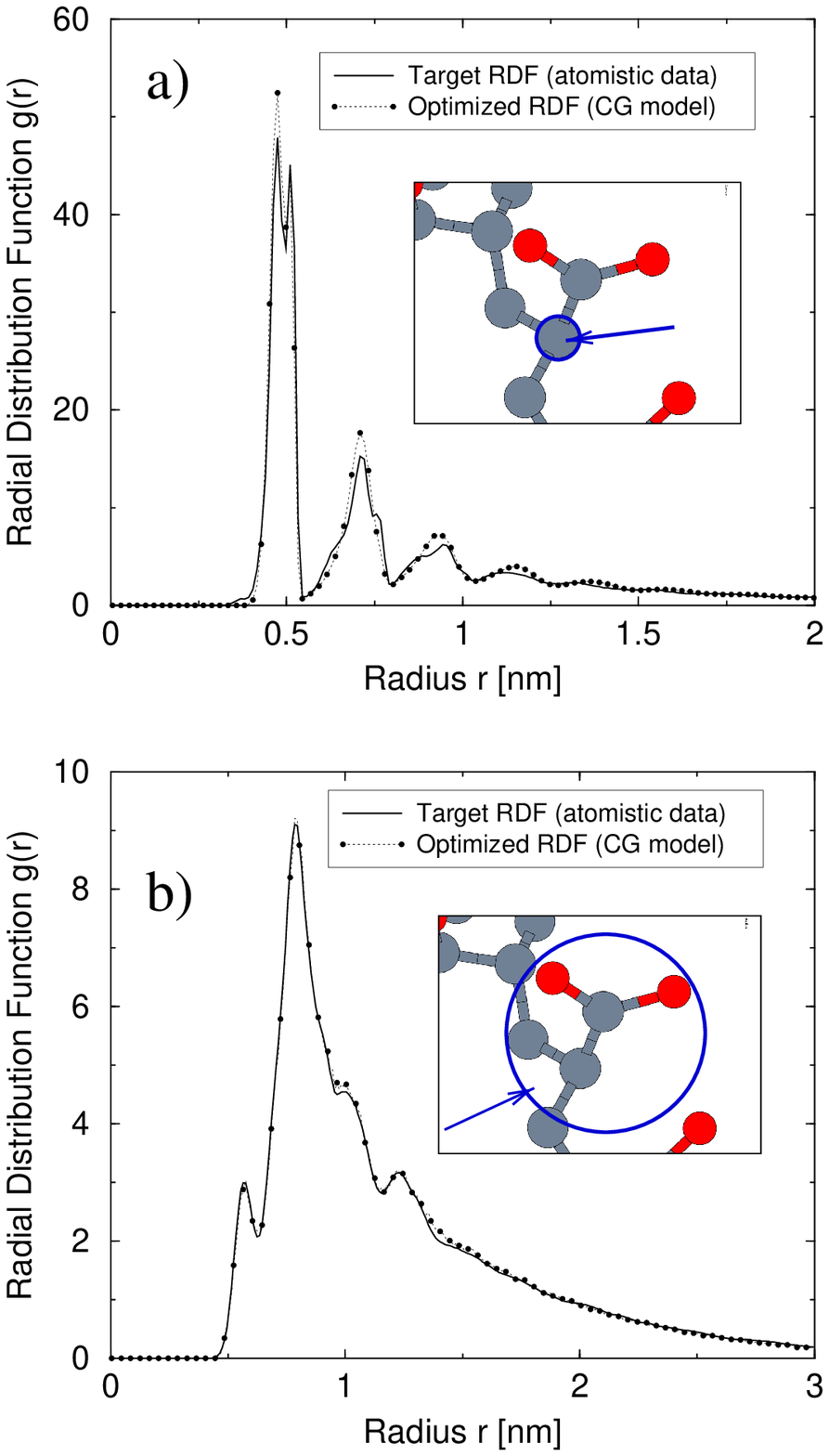}\]
\caption{}\label{fig:rdfs}
\end{figure*}

\begin{figure*}
 \[ \includegraphics[width=\GlobalFigScale\linewidth]{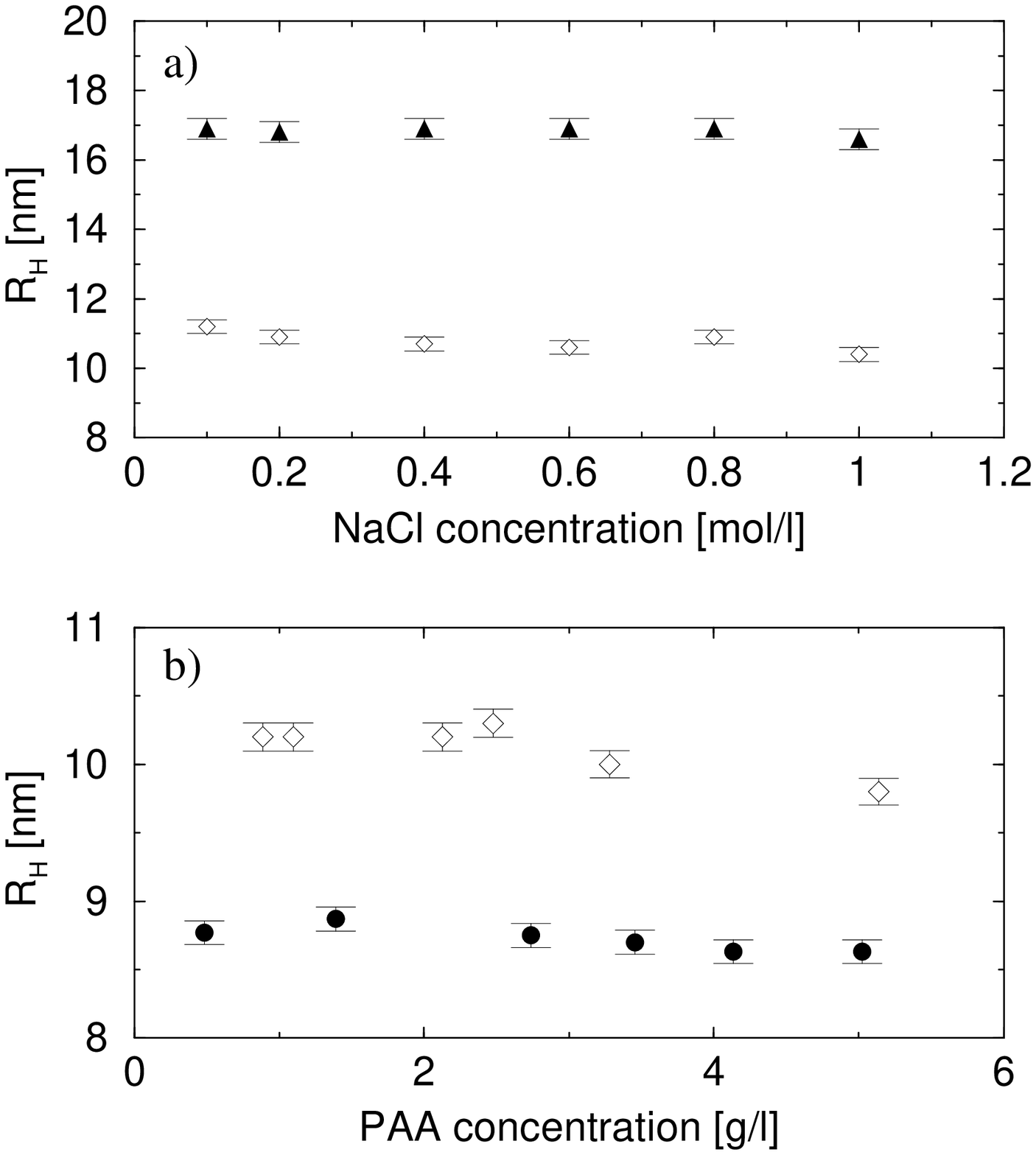}\]
\caption{}\label{fig:exp_checks}
\end{figure*}

\begin{figure*}
 \[ \includegraphics[width=\GlobalFigScale\linewidth]{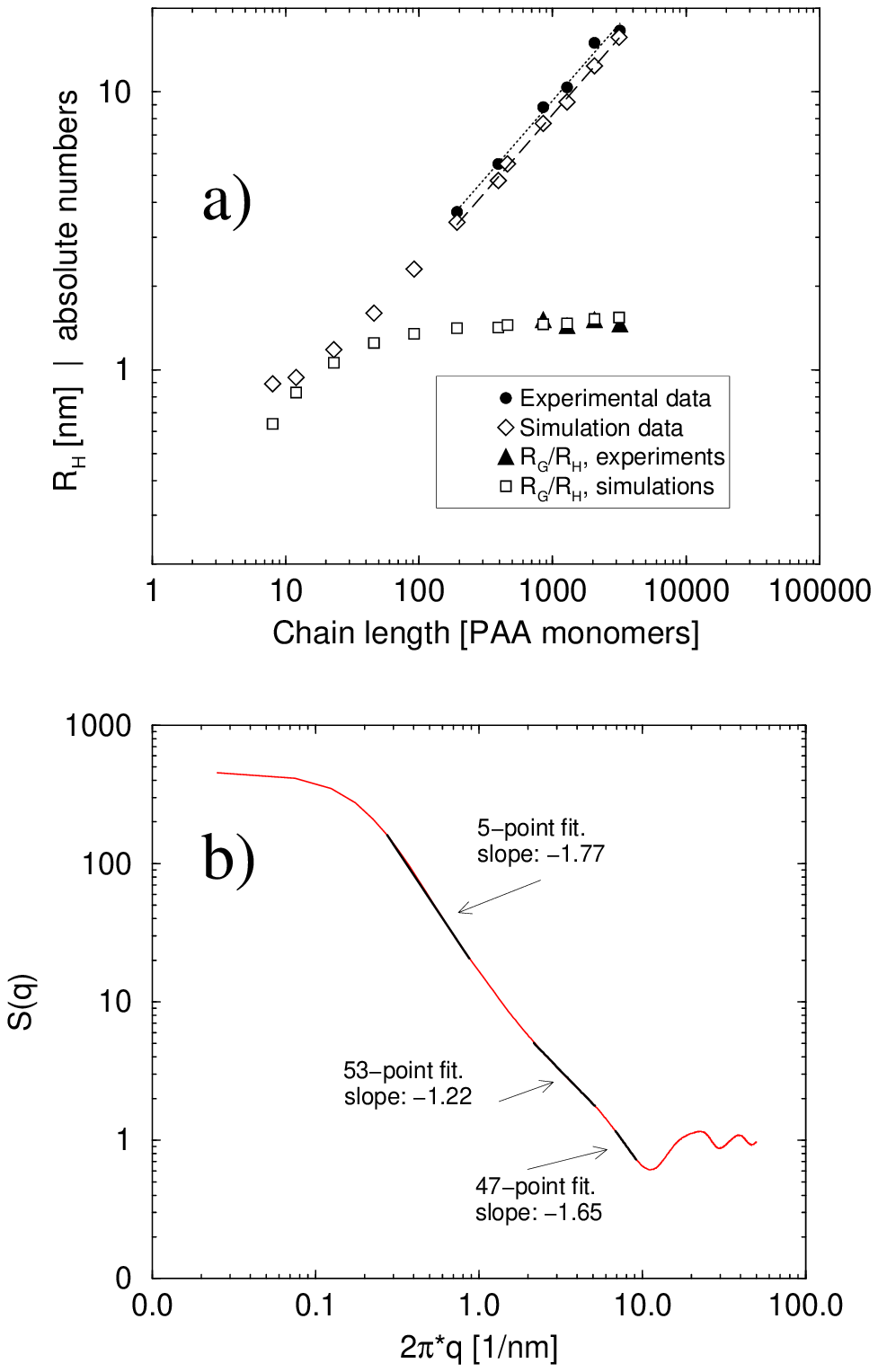}\]
\caption{}\label{fig:scaling}
\end{figure*}

\caplist{
\large{\bf{List of Figures}} \normalsize

Figure \ref{fig:rdfs}: 
Radial distribution functions (RDFs) for coarse-grained
poly (acrylic acid). The monomers are located (a) at the CH1 backbone carbon
(Potential $II$) and (b) at the center of mass of the atomistic repeat unit
(Potential $III$). In both cases, the RDF after optimization almost fully
coincide with the target RDFs from atomistic simulations.

Figure \ref{fig:exp_checks}:
  Hydrodynamic radius for various poly (acrylic acid)
samples as measured by light scattering experiments. (a) Dependence on the
NaCl concentration at constant polymer concentration $1$ g/L (Triangles:
$M_{W}=194300$ g/mol, Diamonds: $M_{W}=119800$ g/mol). (b) Dependence on the
polymer concentration in $1$ M NaCl (Diamonds: $M_{W}=119800$ g/mol, Circles:
$M_{W}=81800$ g/mol).

Figure \ref{fig:scaling}: 
  (a) Scaling behaviour of poly (acrylic acid) as
measured by light scattering experiments (solid symbols) and computer
simulations (open symbols) with coarse-grained model $I$. The hydrodynamic
radius $R_H$ as well as the ratio $R_G/R_H$ (with $R_G$ being the radius of
gyration) is shown. (b) Static structure factor for a $460$-mer, determined by
computer simulations (model $I$). Three different scaling regimes can be seen.

}

\newpage

\end{document}